\RequirePackage{plautopatch} 
\documentclass[12pt]{article}
\usepackage[utf8]{inputenc}

\usepackage[dvipdfmx]{graphicx}
\usepackage{url}
\usepackage{listings,jvlisting}
\usepackage{amssymb}
\usepackage{tabularx}
\usepackage{amsmath}
\usepackage{mdframed}
\usepackage{graphicx}
\usepackage{setspace}
\usepackage{here}
\usepackage{authblk}
\usepackage{mathtools}
\usepackage{amsthm}
\usepackage{algorithmic}
\usepackage{algorithm}
\usepackage{arydshln}

\def\be{{\beta}}

\def\ep{{\varepsilon}}
\def\la{{\lambda}}

\def\si{{\sigma}}

\def\bbe{{\text{\boldmath $\beta$}}}
\def\bga{{\text{\boldmath $\gamma$}}}

\def\bth{{\text{\boldmath $\theta$}}}

\def\bbeh{{\hat \bbe}}

\def\sih{{\hat \si}}

\def\bbeh{{\widehat \bbe}}

\def\bthh{{\widehat \bth}}

\def\De{{\Delta}}

\def\Det{{\widetilde \De}}

\def\y{{\text{\boldmath $y$}}}

\def\C{{\text{\boldmath $C$}}}

\def\M{{\text{\boldmath $M$}}}

\def\R{{\text{\boldmath $R$}}}

\def\X{{\text{\boldmath $X$}}}

\def\Sh{{\widehat S}}

\def\Nc{{\cal N}}

\begin{document}
\title{Variable Selection using Inverse Survival Probability Weighting}
\author[1,2]{Masahiro Kojima\footnote{Address: Biometrics Department, R\&D Division, Kyowa Kirin Co., Ltd.,
Otemachi Financial City Grand Cube, 1-9-2 Otemachi, Chiyoda-ku, Tokyo 100-004, Japan. Tel: +81-3-5205-7200 \quad
E-Mail: masahiro.kojima.tk@kyowakirin.com}}
\affil[1]{Kyowa Kirin Co., Ltd}
\affil[2]{The Institute of Statistical Mathematics}
\maketitle

\abstract{\noindent
In this paper, we propose two variable selection methods for adjusting the censoring information for survival times, such as the restricted mean survival time. To adjust for the influence of censoring, we consider an inverse survival probability weighting (ISPW) for subjects with events. We derive a least absolute shrinkage and selection operator (lasso)-type variable selection method, which considers an inverse weighting for of the squared losses, and an information criterion-type variable selection method, which applies an inverse weighting of the survival probability to the power of each density function in the likelihood function. We prove the consistency of the ISPW lasso estimator and the maximum ISPW likelihood estimator. The performance of the ISPW lasso and ISPW information criterion are evaluated via a simulation study with six scenarios, and then their variable selection ability is demonstrated using data from two clinical studies. The results confirm that ISPW lasso and the ISPW likelihood function produce good estimation accuracy and consistent variable selection. We conclude that our two proposed methods are useful variable selection tools for adjusting the censoring information for survival time analyses.
}
\par\vspace{4mm}

\section{Introduction}\label{sec1}
The restricted mean survival time (RMST) is an important indicator that can be subjected to robust analysis~\cite{huang2018comparison,royston2013restricted}. For such analysis, covariate adjustment provides a means of reducing the confounding bias. Because the analysis of RMST requires the adjustment of the censored subjects, a simple regression model cannot be applied for covariate adjustment. Several covariate adjustment methods for RMST have been proposed~\cite{karrison1987restricted,zucker1998restricted,andersen2004regression,tian2014predicting,ambrogi2022analyzing}. Recently, Andersen et al.~\cite{andersen2004regression} and Tian et al.~\cite{tian2014predicting} reported covariate adjustment methods that directly model the survival time, avoiding the estimation of nonparametric hazards. Andersen's method uses the leave-one-out technique in the analysis of RMST for censored subjects. That is, the pseudo-survival time of all subjects, including the censored subjects, is generated by the leave-one-out method. The generalized regression model can also calculate the RMST adjusted for the covariates. Tian's method analyzes the regression model through inverse survival probability weighting (ISPW) equations for non-censored patients, allowing the censored information to be adjusted. Hanada et al.~\cite{Hanada2022} confirmed that Tian's adjustment method is more stable than that of Andersen et al. In this paper, we focus on Tian's adjustment method. Tian et al.~\cite{tian2014predicting} proposed a cross-validation method for variable selection. However, applying cross-validation to all candidate variable combinations is a time-consuming task.

This paper proposes two novel variable selection methods, one based on the ISPW least absolute shrinkage and selection operator (lasso)~\cite{tibshirani1996regression} and the other based on the ISPW information criterion. For the ISPW lasso method, we propose a minimization problem with an $L_1$ penalty, following Tian's estimation equation. The feature of our proposal is that ISPW is applied to the squared losses. The consistency of the ISPW lasso estimator with the true parameter value is verified. For the information criterion-based method, we define the ISPW likelihood function, which returns the inverse survival probability given the power of the density function of each subject. We develop the ISPW information criterion based on Akaike's information criterion (AIC)~\cite{akaike1998information}. The consistency of the maximum likelihood estimator with the true parameter value is verified. We present a practical example for each variable selection method. The performance of the two variable selection methods is analyzed via computer simulations and data from actual clinical trials.

This paper is organized as follows. Section 2 introduces the methodology behind the proposed ISPW lasso and information criterion approaches. We also present practical example of variable selection. Section 3 describes the setting and results of the computer simulations, before Section 4 presents the results for two actual clinical trials. We conclude this paper with a discussion in Section 5. The R program files used to analyze the simulation results and actual clinical trials are included in the Supplemental Material.

\section{Methods}\label{sec2}
We assume that the survival time variable is $T$ and the right censored variable is $C$, which is independent of $T$. The observed survival time is $R=\min(T,C)$ with censoring indicator $\De=I(T \le C)$. When the restricted time is $\tau$, the observed survival time until $\tau$ is $Y=$min$(R,\tau)$. The covariate $\X$ is a $q$-dimensional vector. The parameter for the covariate is a $q$-dimensional vector $\beta$. The number of subjects is $n$. For subject $i$, the observation is $(Y_i,\De_i,\X_i)$.

We consider two variable selection methods. One is based on lasso and the other is based on an information criterion.

\subsection{Variable selection based on lasso}

We use the lasso shrinkage effects to select variables. Tian et al.~\cite{tian2014predicting} proposed a weighted estimating equation for adjusting the covariate. With reference to this estimating equation, we define a penalized least-squares method with an $L_1$ penalty. The estimator of $\bbe$ can be derived as
\begin{align}
    \underset{\bbe}{\mbox{argmin}}\left\{\frac{1}{N}\sum^n_{i=1}\frac{\Det_i}{\Sh(Y_i)}(h(Y_i)-\X_i^T\bbe)^2+\la||\bbe||_1\right\},
\end{align}
where $\frac{\Det_i}{\Sh(Y_i)}$ is the ISPW, $\Det_i=I(Y_i\leq C_i)$, $\Sh(Y_i)$ is the Kaplan-Meier estimator, $||\cdot||_1$ is the $L_1$ norm, $h(\cdot)$ is an identity function or a log function, and $\la$ is a tuning parameter. One feature of ISPW is that the censored subjects are not used in the analysis. In addition, as the number of subjects at risk decreases over time, the inverse survival probability is used to equalize the information. Because the survival function also reflects censor information, the censor information can be adjusted accordingly.

We now prove the consistency of the estimator $\bbeh$. We state the following assumption.

\bigskip
\noindent
{\bf Assumption 1.} {\it 
\begin{align}
    \frac{1}{n}\sum^n_{i=1}\frac{\Det_i}{S(Y_i)}\X_i\X_i^T\rightarrow \C,
\end{align}
where $\C$ is a nonsingular matrix.
}

Fu and Knight~\cite{fu2000asymptotics} showed that $\bbeh$ converges in probability to argmin$(Z(\bga))$, where
\begin{align}
    Z(\bga)=(\bga-\bbe)^T\C(\bga-\bbe)+\la||\bbe||_1
\end{align}
and argmin(Z(\bga))=\bbe. Thus, $\bbeh$ is consistent with $\bbe$.

\subsection{Model selection based on ISPW information criterion}
We now define the ISPW likelihood function. We assume that $T$ follows a log-normal distribution with parameters $\X_i^T\bbe$ and $\si^2$, a Weibull distribution with scale parameter $\exp(\X_i^T\bbe)$ and shape parameter $\frac{1}{\si}$, or a log-logistic distribution with scale parameter $\exp(\X_i^T\bbe)$ and shape parameter $\frac{1}{\si}$.

The survival time model is 
\begin{align}
    \log(Y_i)=\X_i^T\bbe+\si\ep_i,
\end{align}
where $\ep_i$ is the normal distribution $\Nc(0,1)$ when $T$ follows a log-normal distribution, the Gumbel distribution $Gumbel(0,-1)$ when $T$ follows a Weibull distribution, or the logistic distribution $Logistic(0,-1)$ when $T$ follows a log-logistic distribution. $\X_i$ is a $k$-dimensional covariate vector and $\bbe$ is a $k$-dimensional parameter vector. We assume that $\bth=(\bbe^T,\si^2)^T$.

The inverse probability of censoring weighted likelihood function is expressed as
\begin{align}
    L(\bth,\si^2)=\prod^n_{i=1}f(Y_i|X_i,\bth)^{\frac{\Det_i}{\Sh(Y_i)}},
\end{align}
where $\Det_i=I(Y_i\leq C_i)$, $f(\cdot)$ is a density function, and $\Sh(\cdot)>0$ is the estimated survival probability. The log-likelihood is
\begin{align}
l(\bth)=\log(L(\bth))=\sum^n_{i=1}\frac{\Det_i}{\Sh(Y_i)}\log(f(Y_i|X_i,\bth)).    
\end{align}
The maximum ISPW likelihood estimator $\bthh$ can be obtained by solving the estimating equation 
\begin{align}
    \frac{\partial l(\bth)}{\partial \bth}=0.
\end{align}

We now prove the consistency of $\bthh$.
We assume that $\y=(y_1,y_2,\ldots,y_n)^T$ and $\X=(\X_1,\X_2,\ldots,\X_n)$.

\bigskip
\noindent
{\bf Assumption 2.} {\it Consider an $n\times n$ diagonal matrix $\M_g=\mbox{diag}(g(Y_1,\X_1),g(Y_2,\X_2),\ldots,g(Y_n,\X_n))$, where $g(Y_i,\X_i)\in\R$. $\M_g\X$ is bounded and $\X^T\M_g\X$ is positive-definite. $\X^T\M_g\X/n$ converges to a positive-definite matrix, which implies that $\M_g\X/\sqrt{n}$ converges to its bound.}

\noindent
{\bf Assumption 3.} {\it By using the central limit theorem, for the normal distribution $\ep_i\sim\Nc(0,1)$, $n\left(\mbox{E}[\ep_i^2]-\frac{1}{n}\sum_{i=1}^n\ep_i^2\right)=n\left(1-\frac{1}{n}\sum_{i=1}^n\ep_i^2\right)=nO_p\left(\frac{1}{\sqrt{n}}\right)=O_p(\sqrt{n})$. For the Gumbel distribution $\ep_i\sim Gumbel(0,-1)$, $n\left(\mbox{E}[\ep_i\exp(\ep_i)-\ep_i]-\frac{1}{n}\sum_{i=1}^n(\ep_i\exp(\ep_i)-\ep_i)\right)=n\left(1-\frac{1}{n}\sum_{i=1}^n(\ep_i\exp(\ep_i)-\ep_i)\right)=O_p(\sqrt{n})$. For the logistic distribution $\ep_i\sim Logistic(0,-1)$, $n\left(\mbox{E}\left[\frac{\ep_i}{1+\exp(\ep_i)}\right]-\frac{1}{n}\sum_{i=1}^n\left(\frac{\ep_i}{1+\exp(\ep_i)}\right)\right)=n\left(-\frac{1}{2}-\frac{1}{n}\sum_{i=1}^n\left(\frac{\ep_i}{1+\exp(\ep_i)}\right)\right)=O_p(\sqrt{n})$. The expectations for the Gumbel and logistic distributions are derived in the appendix.}

\bigskip
\noindent
{\bf Theorem} {\it Under Assumptions 2 and 3, and when the number of censored subjects $n_c$ satisfies $0\leq\frac{n_c}{n}<1$, $\bthh$ is consistent with $\bth$. }

\begin{proof} The Taylor expansion of $L(\bthh)$ around $\bth$ is
\begin{align}
   \frac{\partial L(\bthh)}{\partial \bth}=\frac{\partial L(\bth)}{\partial \bth}+\frac{\partial^2 L(\bth)}{\partial \bth\partial \bth^T}(\bthh-\bth)+O_p(1)=0.
\end{align}

From Assumptions 2 and 3, 
\begin{align}
   \bthh&=\bth-\left(\frac{\partial^2 L(\bth)}{\partial \bth\partial \bth^T}\right)^{-1}\frac{\partial L(\bth)}{\partial \bth}+O_p\left(\frac{1}{n}\right)\\
   &=\bth+O_p\left(\frac{1}{\sqrt{n}}\right).
\end{align}
Details of the order of each likelihood function are given in the appendix.
\end{proof}

Because the asymptotic behavior of the ISPW likelihood function is no different from that of the standard likelihood function, we develop an ISPW information criterion that has the same form as the AIC:
\begin{align}
    -2l(\bthh)+2k.
\end{align}

\subsubsection{Practical example of ISPW lasso and AIC}
Practical examples of ISPW lasso and ISPW AIC are now presented to demonstrate the variable selection process. We prepare the example dataset listed in Table \ref{Table1}. The restricted time is $100$.
\begin{table}[H]
  \begin{center}
\caption{Example dataset\label{Table1}}
\begin{tabular}{|c|c|c|c|c|c|c|}\hline
ID& ST & Censor & SP & Group & Age & Sex\\ \hline
1& 20 & No & 0.92 & Control & 70 & F\\
2& 20 & Yes & - & Treat & 60 & F\\
3& 30 & Yes & -  & Control & 60 & M\\
4& 40 & No & 0.71 & Treat & 80 & M \\
5& 40 & No & 0.71 & Control & 60 & F\\
6& 50 & No & 0.61 & Control & 80 & M\\
7& 60 & Yes & -  & Treat & 70 & F\\
8& 80 & No & 0.49 & Treat & 70 & M\\
9& 80 & Yes & -  & Control & 70 & F\\
10&90 & No & 0.33  & Control & 60 & M\\
11&100 & Yes & -  & Treat & 60 & F\\
12&100 & Yes & -  & Treat & 60 & M\\\hline
\end{tabular}
\\
      \footnotesize{ST: Survival time, SP: Survival probability for all subjects with event}
  \end{center}
\end{table}

We assume that $X_{i1}$ is the intercept term, $X_{i2}$ is the treatment group (1: treatment, 0: control), $X_{i3}$ is age, $X_{i4}$ is sex (1: male, 0: female), and $\X_i=(X_{i1},X_{i2},X_{i3},X_{i4})^T$ for subject $i$. The estimators $\bbeh$ of Tian's method are $(6.52,0.37,-0.04,0.37)$. The p-values for each estimator are $0.000, 0.020, 0.001, 0.016$. ISPW lasso considers the minimization problem for the following equation:

\begin{align}
    \frac{1}{12}\biggl(&\frac{1}{0.92}(\log(20)-\X_1^T\bbe)^2+\frac{1}{0.71}(\log(40)-\X_4^T\bbe)^2+\frac{1}{0.71}(\log(40)-\X_5^T\bbe)^2\nonumber\\
    &\hspace{0.5cm}+\frac{1}{0.61}(\log(50)-\X_6^T\bbe)^2+\frac{1}{0.49}(\log(80)-\X_8^T\bbe)^2+\frac{1}{0.33}(\log(90)-\X_9^T\bbe)^2\biggr)\nonumber\\
    &\hspace{1cm}+0.1|\be_1|+0.1|\be_2|+0.1|\be_3|+0.1|\be_4|.
\end{align}

For a fixed tuning parameter $\la=0.10$, the estimator $\bbeh$ of ISPW lasso is $(4.91,0.00,-0.02,0.71)$. With $\la=0.05$ after cross-validation tuning, the estimator $\bbeh$ of ISPW lasso is $(5.36,0.00,-0.03,0.87)$.

The maximum ISPW likelihood estimator is calculated by
\begin{align}
    l(\bth)=\frac{1}{12}\biggl(&\frac{1}{0.92}\log(f(20|X_1,\bth))+\frac{1}{0.71}\log(f(40|X_4,\bth))+\frac{1}{0.71}\log(f(40|X_5,\bth))\nonumber\\
    &\hspace{0.5cm}+\frac{1}{0.61}\log(f(50|X_6,\bth))+\frac{1}{0.49}\log(f(80|X_8,\bth))+\frac{1}{0.33}\log(f(90|X_9,\bth))\biggr).
\end{align}

The maximum ISPW likelihood estimators and ISPW AICs are listed in Table \ref{Table2}.
\begin{table}[H]
  \begin{center}
\caption{Maximum ISPW likelihood estimators and ISPW AICs of example dataset\label{Table2}}
\begin{tabular}{|c|c|c|c|}\hline
Distribution& Variables & MLEs & AICs\\ \hline
Log-normal & (intercept, treatment group, age, sex, error) & (5.93, 0.09, -0.04, 1.01, 0.02) & 81.88\\
Log-normal & (intercept, treatment group, sex, error) & (3.39, -0.20, 0.91, 0.10) & 99.26\\
Log-normal & (intercept, age, sex, error) & (5.81, -0.04, 1.04, 0.02) & \textcolor{red}{\bf 80.61}\\
Log-normal & (intercept, sex, error) & (3.39, 0.83, 0.11) & 97.99\\\hdashline
Weibull & (intercept, treatment group, age, sex, error) & (4.64, 0.08, -0.01, 0.53, 0.19) & 96.13\\
Weibull & (intercept, treatment group, sex, error) & (3.57, -0.14, 0.85, 0.23) & 97.27\\
Weibull & (intercept, age, sex, error) & (2.68, 0.01, 0.82, 0.31) & 104.16\\
Weibull & (intercept, sex, error) & (3.56, 0.80, 0.24) & \textcolor{red}{95.97}\\\hdashline
Log-logistic & (intercept, treatment group, age, sex, error) & (7.12, 0.24, -0.06, 1.12, 0.12) & 91.08\\
Log-logistic & (intercept, treatment group, sex, error) & (3.41, -0.19, 0.91, 0.20) & 100.65\\
Log-logistic & (intercept, age, sex, error) & (5.76, -0.04, 1.01, 0.08) & \textcolor{red}{81.93}\\
Log-logistic & (intercept, sex, error) & (3.41, 0.84, 0.20) & 99.20\\\hline
\end{tabular}
\\
      \footnotesize{ST: Survival time, SP: Survival probability for all subjects. Bold denotes the smallest AIC. Red denotes the smallest AIC in each distribution.}
  \end{center}
\end{table}
     
The minimum ISPW AIC is $80.61$. In the case with the smallest AIC, the maximum ISPW likelihood estimator $\bbeh$ on the log-normal distribution is $(5.81,0.00,-0.04,1.04)$ and $\sih$ is $0.02$. The variables chosen by both ISPW lasso and ISPW AIC are the intercept, age, and sex. The treatment group, which is the variable with the highest p-value in Tian's method, is not selected.

\section{Simulation study}
\subsection{Simulation configuration}
We performed a simulation study to evaluate the performance of ISPW lasso and ISPW AIC. The evaluation indicators are the mean squared error (MSE) and the percentage of cases in which the correct combination of variables is selected. We generated the survival data $\y$ based on the simulation settings in Table \ref{Table3}. A total of 10,000 simulations were performed for each scenario. The lasso tuning parameter $\la$ was fixed to 0.1. The case in which this parameter is tuned by cross-validation is considered in the appendix.
\begin{table}[H]
  \begin{center}
\caption{Parameter values of simulation scenarios \label{Table3}}
\begin{tabular}{|c|c|c|c|c|c|c|c|c|}
\hline
  & & \multicolumn{5}{c|}{True parameter values} & \multicolumn{2}{c|}{\%Censored} \\
Scenario & True generating model & \multicolumn{1}{c}{$\be_0$} & \multicolumn{1}{c}{$\be_1$} & \multicolumn{1}{c}{$\be_2$} & \multicolumn{1}{c}{$\si$} & \multicolumn{1}{c|}{$\ep$} & \multicolumn{1}{c}{trt$=0$} & trt$=1$ \\ \hline
1 & $\exp(\be_0+\mbox{trt}\be_1+X_1\be_2+\si\ep)$ & 1 & 1 & 1& 1 & Normal & 0.1 & 0.1\\
2 & $\exp(\be_0+\mbox{trt}\be_1+X_1\be_2+\si\ep)$ & 1 & 1 & 1& 1 & Normal & 0.1 & 0.3\\
3 & $\exp(\be_0+\mbox{trt}\be_1+X_1\be_2+\si\ep)$ & 1 & 1 & 1& 1 & Gumbel & 0.1 & 0.1\\
4 & $\exp(\be_0+\mbox{trt}\be_1+X_1\be_2+\si\ep)$ & 1 & 1 & 1& 1 & Gumbel & 0.1 & 0.3\\
5 & $\exp(\be_0+\mbox{trt}\be_1+X_1\be_2+\si\ep)$ & 1 & 1 & 1& 1 & Logistic & 0.1 & 0.1\\
6 & $\exp(\be_0+\mbox{trt}\be_1+X_1\be_2+\si\ep)$ & 1 & 1 & 1& 1 & Logistic & 0.1 & 0.3\\\hline
\end{tabular}
\\
      \footnotesize{$X_1\sim\Nc(1,1)$, trt was randomly assigned in a 1:1 ratio to 1 and 0.}
\end{center}
\end{table}

\subsection{Simulation results}
The MSEs for each method are presented in Table \ref{Table4}. There is little difference in the MSEs of the lasso- and likelihood estimator-based methods compared with Tian's method.
\begin{table}[H]
  \begin{center}\label{Table4}
\caption{MSE results }
\begin{tabular}{|c|c|c|c|c|c|c|}\hline
&\multicolumn{3}{c|}{Sample size=200}&\multicolumn{3}{c|}{Sample size=1000}\\\hline
&Tian&lasso&likelihood&Tian&lasso&likelihood\\\hline
\multicolumn{7}{|l|}{Scenario 1}\\\hline
$\be_0$&0.063&0.021 &0.055&0.021 &0.007 & 0.006\\
$\be_1$&0.137&0.223 &0.129&0.075 &0.204 & 0.054\\
$\be_2$&0.152&0.127 &0.155&0.057 &0.122 & 0.051\\
$\si$ &-& - & 0.066 &-& - &0.053\\\hline
\multicolumn{7}{|l|}{Scenario 2}\\\hline
$\be_0$&0.101&0.023 &0.023 &0.093&0.005 & 0.004\\
$\be_1$&0.181&0.217 &0.071 &0.173&0.194 & 0.049\\
$\be_2$&0.171&0.127 &0.056 &0.175&0.121 & 0.051\\
$\si$ &-& - & 0.068 &-& - &0.053\\\hline
\multicolumn{7}{|l|}{Scenario 3}\\\hline
$\be_0$&0.326&0.045 & 0.252 &0.321&0.032 & 0.232\\
$\be_1$&0.284&0.336 & 0.370 &0.277&0.314 & 0.355\\
$\be_2$&0.294&0.211 & 0.337 &0.297&0.205 & 0.347\\
$\si$ &-& - & 0.086 &-& - &0.052\\\hline
\multicolumn{7}{|l|}{Scenario 4}\\\hline
$\be_0$&0.432&0.043 & 0.235 &0.427&0.026 & 0.214\\
$\be_1$&0.376&0.329 & 0.352 &0.369&0.300 & 0.341\\
$\be_2$&0.324&0.209 & 0.325 &0.327&0.201 & 0.335\\
$\si$ &-& - & 0.065 &-& - &0.053\\\hline
\multicolumn{7}{|l|}{Scenario 5}\\\hline
$\be_0$&0.188&0.112 & 0.064 &0.175&0.080 &0.032 \\
$\be_1$&0.309&0.388 & 0.197 &0.296&0.351 &0.158 \\
$\be_2$&0.322&0.244 & 0.162 &0.310&0.234 &0.150 \\
$\si$ &-& - & 0.074 &-& - &0.068\\\hline
\multicolumn{7}{|l|}{Scenario 6}\\\hline
$\be_0$&0.294&0.153 & 0.098 &0.282&0.112 &0.056 \\
$\be_1$&0.421&0.380 & 0.197 &0.406&0.341 &0.148 \\
$\be_2$&0.340&0.247 & 0.161 &0.338&0.233 &0.147 \\
$\si$ &-& - & 0.073 &-& - &0.065\\\hline
\end{tabular}
\end{center}
\end{table}

We now consider the percentage of cases in which the correct combination of variables is selected. We denote $\be_0+\mbox{trt}\be_1+X_1\be_2$ as candidate variable combination 1 (C1), $\be_0+\mbox{trt}\be_1+X_1\be_2$ as combination 2 (C2), and $\be_0+\mbox{trt}\be_1+X_1\be_2+X_2\be_3$ as combination 3 (C3), where $X_2$ follows the normal distribution $\Nc(-1,1)$. For lasso, any other combination is denoted as combination 4 (C4). Combination 2 is the correct combination. The results are presented in Table \ref{Table5}. With the lasso-based method, the percentage of correct combinations increases from 60\% to around 80\%--90\% as the sample size increases. The AIC remains stable at 60\%--80\% for all distributions.
\begin{table}[H]
  \begin{center}
\caption{Percentage of times the correct variable combination is selected\label{Table5}}
\begin{tabular}{|c|c|c|c|c|}\hline
&\multicolumn{2}{c|}{Sample size=200}&\multicolumn{2}{c|}{Sample size=1000}\\\hline
&lasso&likelihood&lasso&likelihood\\\hline
\multicolumn{5}{|l|}{Scenario 1}\\\hline
C1& 0.0 & 0.0 & 0.0 & 0.0\\
\bf C2&\bf 80.3 &\bf 60.1 &\bf 99.4 &\bf 60.5\\
C3& 19.7 & 39.9& 0.6 & 39.5\\
C4& 0.0 & - & 0.0 & - \\\hline
\multicolumn{5}{|l|}{Scenario 2}\\\hline
C1& 0.0 & 0.0 & 0.0 & 0.0\\
\bf C2&\bf 78.8&\bf 63.4 &\bf 99.3 &\bf 63.6\\
C3& 21.2& 36.6 & 0.7 & 36.4\\
C4& 0.1 & - & 0.0 & - \\\hline
\multicolumn{5}{|l|}{Scenario 3}\\\hline
C1& 0.0 & 0.9 & 0.0 & 0.0\\
\bf C2&\bf 79.9 &\bf 75.3 &\bf 99.7 &\bf 71.9\\
C3& 19.8 & 23.8 & 0.3 & 28.1\\
C4& 0.4 & - & 0.0 & - \\\hline
\multicolumn{5}{|l|}{Scenario 4}\\\hline
C1& 0.0 & 0.9 & 0.0 & 0.0\\
\bf C2&\bf 77.0 &\bf 81.1 &\bf 99.3 &\bf 81.7\\
C3& 22.6 & 18.0& 0.7 & 18.3\\
C4& 0.5 & - & 0.0 & - \\\hline
\multicolumn{5}{|l|}{Scenario 5}\\\hline
C1& 0.0 & 0.0 & 0.0 & 0.0\\
\bf C2&\bf 59.0 &\bf 64.5 &\bf 95.1 &\bf 65.3\\
C3& 37.0 & 35.5& 4.9 & 34.7\\
C4& 4.0 & - & 0.0 & - \\\hline
\multicolumn{5}{|l|}{Scenario 6}\\\hline
C1& 0.0 & 0.0 & 0.0 & 0.0\\
\bf C2&\bf 56.9 &\bf 67.7 &\bf 93.7 &\bf 68.3\\
C3& 39.0 & 32.3 & 6.3 & 31.7\\
C4& 4.1 & - & 0.0 & - \\\hline
\end{tabular}
\\
      \footnotesize{Bold denotes the percentage of correct variable combinations.}
\end{center}
\end{table}

\section{Actual study}
We now demonstrate the performance of the proposed variable selection methods for two actual clinical studies. The first trial evaluated survival times in primary biliary cholangitis (PBC), with more than half of the data censored~\cite{neuberger1985double}. The second trial evaluated survival times in malignant glioma (MG), in which few censored data were observed~\cite{brem1995placebo}. 

\subsection{Analysis for PBC}
\begin{figure}[H]
  \begin{center}
  \includegraphics[width=15cm]{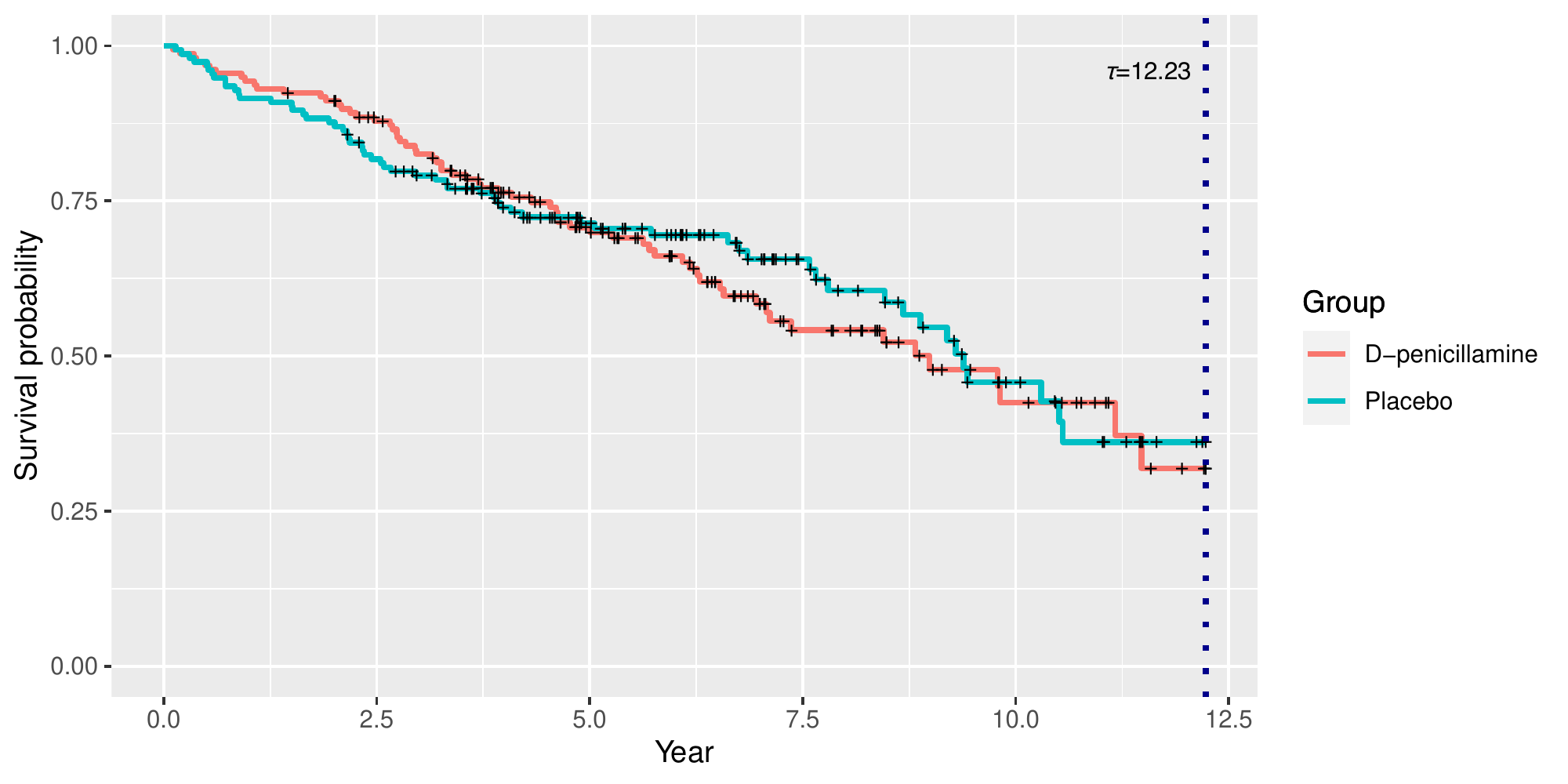}
  \caption{Survival curve of PBC dataset}
  \label{Figure1}
      \footnotesize{$+$ denotes censoring, $\tau$ is the restricted time.}
  \end{center}
\end{figure}

The randomized placebo-controlled trial of PBC evaluated survival time (years) from enrollment to death for the D-penicillamine group versus the placebo group. The number of randomized subjects was 134, with 72 in the D-penicillamine group and 62 in the placebo group. Censoring occurred in 52 subjects in the D-penicillamine group and 46 subjects in the placebo group. The restricted time $\tau$ is 12.23. The survival curve is shown in Figure \ref{Figure1}.

The baseline information of the PBC study concerns edema (0: no edema, 0.5: untreated or successfully treated, 1: edema despite diuretic therapy), serum bilirubin (mg/dL), serum albumin (mg/dL), standardized blood clotting time, and age (years). Table \ref{Table6} presents the results using Tian's adjustment method when the link function is a log function.

\begin{table}[H]
  \begin{center}
\caption{Analysis results given by Tian's method using PBC data\label{Table6}}
\begin{tabular}{|l|l|l|}\hline
Parameter &  Mean & p-value [$95\%$CI]\\ \hline
Intercept & 1.93 & 0.000 [1.01, 2.85]\\
Treatment group & 0.09 & 0.580 [-0.23, 0.41]\\
Edema & -2.03 & 0.000 [-2.66, -1.39]\\
Serum bilirubin & -0.12 & 0.000 [-0.18, -0.06]\\
Serum albumin & 0.02 & 0.830 [-0.18, 0.23]\\
Standardized blood clotting time & -0.01 & 0.760 [-0.09, 0.07]\\
Age & 0.01 & 0.110 [0.00, 0.02]\\
\hline
\end{tabular}
\\
      \footnotesize{SE: Standard Error. $95\%$CI: $95\%$ Confidence Interval.}
  \end{center}
\end{table}

The variable selection results using ISPW lasso and the minimum ISPW AIC for each distribution are given in Table \ref{Table7}.

\begin{table}[H]
  \begin{center}
\caption{Analysis results given by ISPW lasso and ISPW likelihood function\label{Table7}}
\begin{tabular}{|l|c|c|c|c|}\hline
Parameter & lasso & likelihood (LN) & likelihood (W) & likelihood (LL)\\ \hline
Intercept & 1.40 & 1.66 & 1.92 & 1.94\\
Treatment group & -0.02 & - & - & -\\
Edema & -1.76 & -2.31 & -1.71 & -2.08\\
Serum bilirubin & -0.02 & - & - & -0.08\\
Serum albumin & 0.07 & - & - & -\\
Standardized blood clotting time & - & - & - & -\\
Age & - & - & - & -\\
Error & - & 0.61 & 0.53 & 0.39\\\hdashline
AIC & - & 232.7 & \textcolor{red}{\bf 218.2} & 228.5\\
\hline
\end{tabular}
\\
      \footnotesize{LN: Log-normal, W: Weibull, LL: Log-logistic}
  \end{center}
\end{table}

The lasso-based method does not select the standardized blood clotting time or age. Additionally, the estimated results are slightly closer to zero than in Tian's results. For the AIC-based method, the intercept and edema are selected. In the AIC case, the results given by Tian's analysis with the smallest p-value are selected.

\subsection{Analysis for MG}
\begin{figure}[H]
  \begin{center}
  \includegraphics[width=15cm]{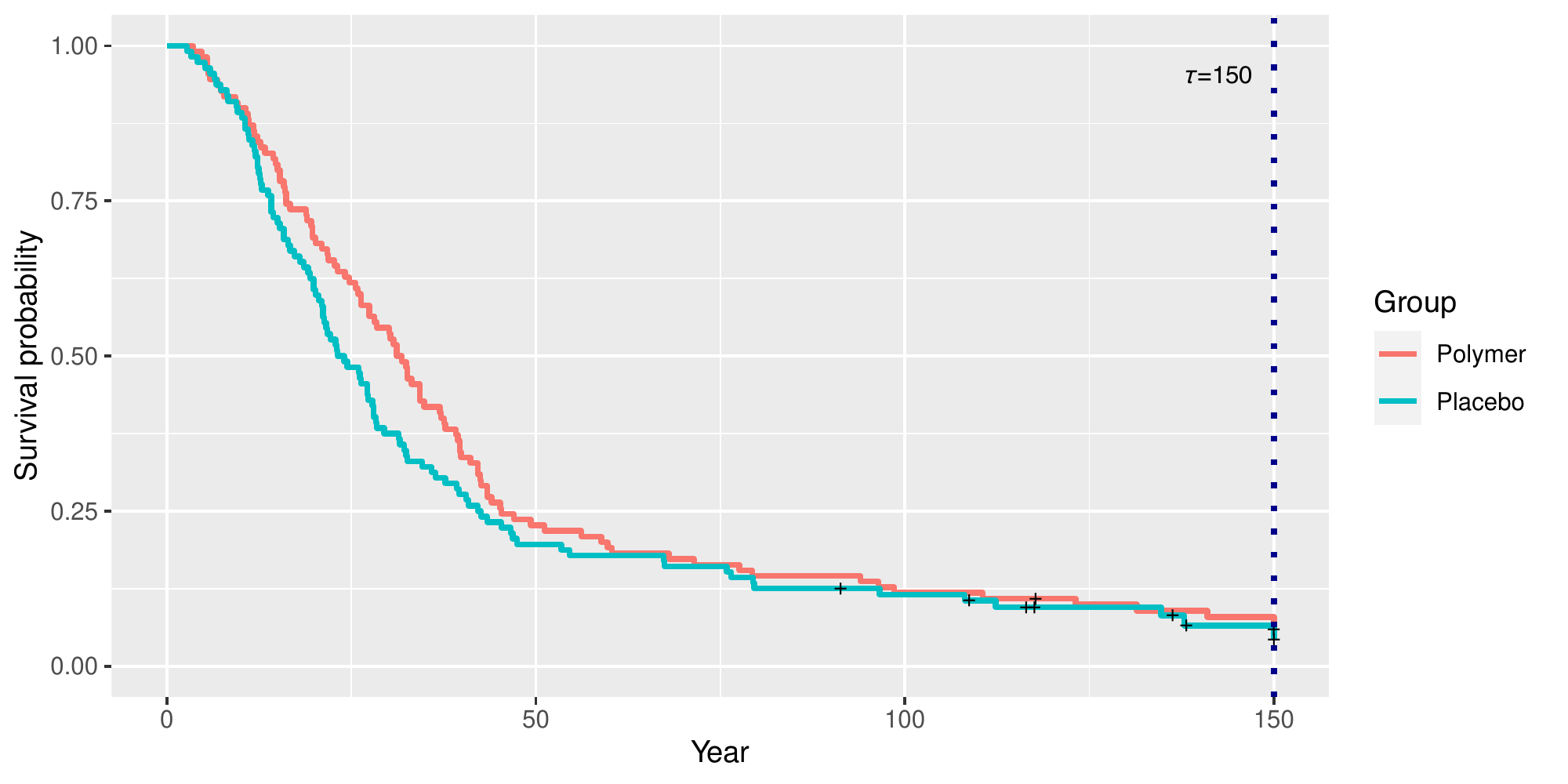}
  \caption{Survival curve of MG dataset}
  \label{Figure2}
      \footnotesize{$+$ denotes censoring, $\tau$ is the restricted time.}
  \end{center}
\end{figure}

The randomized placebo-controlled clinical trial for MG evaluated survival time (weeks) from enrollment to death for the polymer (chemotherapeutic agents incorporated into biodegradable polymers) group and placebo group. The number of randomized subjects was 222, with 110 in the polymer group and 112 in the placebo group. Censoring occurred in seven subjects in the polymer group and eight subjects in the placebo group. The restricted time $\tau$ is 150. The survival curve is shown in Figure \ref{Figure2}.

The baseline information concerns age (years), years since diagnosis, Karnofsky performance score (0: $<$70, 1: $\geq$ 70), race (0: other, 1: white), radiation therapy (0: $<$45 Gy, 1: $\geq$45 Gy), sex (0: female, 1:male), nitro, tumor histopathology at implementation (path) (1: glioblastoma, 2: anaplastic astrocytoma, 3: oligodendroglioma, 4: other), and grade (0: quiescent, 1: active). Table \ref{Table8} presents the results using Tian's adjustment method when the link function is a log function.

\begin{table}[H]
  \begin{center}
\caption{Analysis results given by Tian's methods using MG data\label{Table8}}
\begin{tabular}{|l|l|l|}\hline
Parameter & Mean & p-value [$95\%$CI]\\ \hline
Intercept & 3.74 & 0.000 [2.95, 4.52]\\
Treatment group & 0.37 & 0.001 [0.15, 0.58]\\
Age & -0.02 & 0.001 [-0.02, -0.01]\\
Years of diagnosis & 0.02 & 0.224 [-0.01, 0.05]\\
Karnofsky performance score & 0.37 & 0.003 [0.12, 0.62]\\
Race & -0.21 & 0.240 [-0.57, 0.14]\\
Radiation therapy & 0.35 & 0.010 [0.08, 0.61]\\
Sex & 0.12 & 0.348 [-0.13, 0.36]\\
Nitro & -0.27 & 0.031 [-0.51, -0.02]\\
Path & 0.27 & 0.000 [0.14, 0.40]\\
Grade & -0.32 & 0.115 [-0.73, 0.08]\\
\hline
\end{tabular}
\\
      \footnotesize{SE: Standard Error. CI: $95\%$ Confidence Interval.}
  \end{center}
\end{table}

The variable selection results using the ISPW lasso and minimum ISPW AIC methods for each distribution are listed in Table \ref{Table9}

\begin{table}[H]
  \begin{center}
\caption{Analysis results given by ISPW lasso and ISPW likelihood function using MG data\label{Table9}}
\begin{tabular}{|l|c|c|c|c|}\hline
Parameter & lasso & likelihood (LN) & likelihood (W) & likelihood (LL)\\ \hline
Intercept & 3.78 & 4.47 & 2.40 & 2.64\\
Treatment group & - & 0.17 & 0.11 & 0.19\\
Age & - & -0.01 & 0.02 & 0.00\\
Years of diagnosis & 0.01 & 0.03 & - & -\\
Karnofsky performance score & 0.19 & 0.30 & - & 0.70\\
Race & -0.12 & -0.39 & - & -\\
Radiation therapy & - & 0.48 & - & -\\
Sex & 0.23 & - & - & -\\
Nitro & -0.18 & -0.38 & - & -\\
Path & 0.02 & 0.06 & 0.43 &0.28 \\
Grade & -0.01 & -0.58 & - & -\\
Error & - & 0.42 & 0.83 & 0.38\\\hdashline
AIC & - & \textcolor{red}{\bf 5387.56} & 5804.40 &5582.18\\
\hline
\end{tabular}
\\
      \footnotesize{LN: Log-normal, W: Weibull, LL: Log-logistic}
  \end{center}
\end{table}

The lasso-based method does not select the treatment group, age, and radiation therapy. For AIC, with the log-normal distribution, all variables are selected except sex. The Weibull and log-logistic distributions select almost the same variables, although in the log-logistic case, the Karnofsky performance score is also selected. 

\section{Discussion}
We have proposed novel ISPW lasso and ISPW AIC methods for selecting covariates of RMST. Tian et al.~\cite{tian2014predicting} proposed a covariate adjustment method based on cross-validation that achieves superior performance~\cite{Hanada2022}, but is computationally time-consuming. The ISPW lasso method adjusts for the censored subjects and the reduction in the number of subjects at risk through the squared loss weighted by the inverse survival probability. We have shown that the estimator of ISPW lasso is consistent with that given by Tian's method. For the ISPW AIC method, we defined a likelihood function with inverse weighting of the survival probability given the power of the density function. The intuitive interpretation of the ISPW likelihood function is that we reduce the number of subjects and adjust the censored data, as in ISPW lasso. We proved that the estimator of the ISPW likelihood function is consistent with that given by Tian's method. The ISPW AIC formula simply changes the standard AIC likelihood function to the ISPW likelihood function. Clear practical examples of these two variable selection methods have been presented.

We evaluated the performance of ISPW lasso and ISPW AIC via a simulation study that examined six scenarios. The MSEs of the estimators of ISPW lasso and the ISPW likelihood function were found to be only slightly different from those given by Tian's method. When the survival times followed a log-normal distribution, the ISPW likelihood estimator was more accurate than that of Tian's method and the lasso-based estimator. In terms of variable selection performance, the lasso-based method achieved high accuracy for true variable combinations with the log-normal and Weibull distributions, although the performance of lasso with tuning parameter set by cross-validation was low, at just under 40\%. As a result, most variables were retained. However, the MSE was improved by cross-validation.
The method based on AIC was able to identify the correct combination of variables in about 70\% of cases across all scenarios.

In our analysis of real data from a PBC study, more variables were selected by lasso than by AIC, but the results presented in the appendix confirm that increasing the tuning parameter results similar to that of AIC. For the MG study, the selected variables differed depending on the distribution, with the log-normal distribution having the lowest AIC.

 We confirmed that ISPW lasso and the ISPW likelihood function provide consistent estimation accuracy and that variable selection can be performed without problems. We conclude that our two proposed methods are useful variable selection techniques for adjusting the censoring data for survival time analyses.

\bigskip
\noindent
{\bf Acknowledgements.}
MK would like to thank Associate Professor Hisashi Noma for his encouragement and helpful suggestions.

\bibliography{main.bib} 
\bibliographystyle{unsrt} 

\appendix
\section{Appendix}\label{App}
\subsection{Asymptotic properties of ISPW log-likelihood function of log-normal distribution}
The ISPW log-likelihood function of the log-normal distribution is 
\begin{align}
    l(\bth)=\sum^n_{i=1}\frac{\Det_i}{\Sh(Y_i)}\left(-\frac{1}{2}\log(2y_i^2\pi\si^2)-\frac{1}{2\si^2}(\log(y_i)-\X_i^T\bbe)^2\right).
\end{align}

The first derivative of the likelihood function with respect to $\bbe$ is
\begin{align}
\label{eq:lognfb}
    \frac{\partial l(\bth)}{\partial \bbe}=-\frac{1}{\si^2}\sum^n_{i=1}\frac{\Det_i}{\Sh(Y_i)}(\log(y_i)-\X_i^T\bbe)\X_i.
\end{align}
From Assumption 2, $\frac{\partial l(\bth)}{\partial \bbe}=O_p(\sqrt{n})$.

The second derivative of the likelihood function with respect to $\bbe$ is
\begin{align}
\label{eq:lognfs}
    \frac{\partial^2 l(\bth)}{\partial \bbe\partial \bbe^T}=-\frac{1}{\si^2}\sum^n_{i=1}\frac{\Det_i}{\Sh(Y_i)}\X_i\X_i^T.
\end{align}

From Assumption 2, $\frac{\partial^2 l(\bth)}{\partial \bbe\partial \bbe^T}=O_p(n)$.

The first derivative of the likelihood function with respect to $\si^2$ is
\begin{align}
    \frac{\partial l(\bth)}{\partial \si^2}=-\frac{1}{2\si^2}\sum^n_{i=1}\frac{\Det_i}{\Sh(Y_i)}\left(1-\frac{1}{\si^2}(\log(y_i)-\X_i^T\bbe)^2\right)
\end{align}
From Assumption 2, $\frac{\partial l(\bth)}{\partial \si^2}=O_p(\sqrt{n})$.

The second derivative of the likelihood function with respect to $\si^2$ is
\begin{align}
    \frac{\partial^2 l(\bth)}{(\partial \si^2)^2}=\frac{1}{2\si^4}\sum^n_{i=1}\frac{\Det_i}{\Sh(Y_i)}\left(1-\frac{1}{4\si^2}(\log(y_i)-\X_i^T\bbe)^2\right).
\end{align}
Because the sum of the constants is $n$, $\frac{\partial^2 l(\bth)}{(\partial \si^2)^2}=O_p(n)$.

The derivative of the likelihood function with respect to $\bbe$ and $\si^2$ is
\begin{align}
    \frac{\partial^2 l(\bth)}{\partial \si^2\partial \bbe}=\frac{1}{\si^4}\sum^n_{i=1}\frac{\Det_i}{\Sh(Y_i)}(\log(y_i)-\X_i^T\bbe)\X_i.
\end{align}
From Assumption 2, $\frac{\partial^2 l(\bth)}{\partial \si^2\partial \bbe}=O_p(\sqrt{n})$.

For the log-normal distribution, the formulas for the estimators of the parameters are given by Eqs.~(\ref{eq:lognfb}) and (\ref{eq:lognfs}), and 
\begin{align}
    \bbeh=\sum^n_{i=1}\frac{\Det_i\log(y_i)}{\Sh(Y_i)}\left(\sum^n_{j=1}\frac{\Det_j}{\Sh(Y_j)}\X_j\X_j^T\right)^{-1}\X_i,
\end{align}
\begin{align}
    \sih^2=\frac{\sum^n_{i=1}\frac{\Det_i}{\Sh(Y_i)}(\log(y_i)-\X_i^T\bbeh)^2}{\sum^n_{i=1}\frac{\Det_i}{\Sh(Y_i)}}.
\end{align}

\subsection{Asymptotic properties of ISPW log-likelihood function of Weibull distribution}
The ISPW log-likelihood function of the Weibull distribution is 
\begin{align}
    l(\bth)=\sum^n_{i=1}\frac{\Det_i}{\Sh(Y_i)}\left(-\log(\si)-\frac{1}{\si}\X_i^T\bbe+\left(\frac{1}{\si}-1\right)\log(y_i)-y_i^\frac{1}{\si}\exp\left(-\frac{\X_i^T\bbe}{\si}\right)\right).
\end{align}

The first derivative of the likelihood function with respect to $\bbe$ is
\begin{align}
    \frac{\partial l(\bth)}{\partial \bbe}=-\frac{1}{\si}\sum^n_{i=1}\frac{\Det_i}{\Sh(Y_i)}\left(1-y_i^\frac{1}{\si}\exp\left(-\frac{\X_i^T\bbe}{\si}\right)\right)\X_i.
\end{align}
From Assumption 2, $\frac{\partial l(\bth)}{\partial \bbe}=O_p(\sqrt{n})$.

The second derivative of the likelihood function with respect to $\bbe$ is
\begin{align}
    \frac{\partial l(\bth)}{\partial \bbe\partial \bbe^T}=-\frac{1}{\si^2}\sum^n_{i=1}\frac{\Det_i}{\Sh(Y_i)}y_i^\frac{1}{\si}\exp\left(-\frac{\X_i^T\bbe}{\si}\right)\X_i\X_i^T.
\end{align}
From Assumption 2, $\frac{\partial l(\bth)}{\partial \bbe\partial \bbe^T}=O_p(n)$.

The first derivative of the likelihood function with respect to $\si$ is
\begin{align}
    \frac{\partial l(\bth)}{\partial         \si}=&\sum^n_{i=1}\frac{\Det_i}{\Sh(Y_i)}\left(-\frac{1}{\si}+\frac{1}{\si^2}\X_i^T\bbe-\frac{1}{\si^2}\log(y_i)+\frac{y_i^\frac{1}{\si}\exp\left(-\frac{\X_i^T\bbe}{\si}\right)}{\si^2}(\log(y_i)-\X_i^T\bbe)\right)\nonumber\\
    =&\sum^n_{i=1}\frac{\Det_i}{\Sh(Y_i)}\left(-\frac{1}{\si}+\frac{\ep_i}{\si}+\frac{\ep_i\exp(\ep_i)}{\si}\right).
\end{align}

The expectation of $\ep_i\exp(\ep_i)-\ep_i$ is
\begin{align}
    E[\ep_i\exp(\ep_i)-\ep_i]=&\int^\infty_{-\infty}(\ep_i\exp(\ep_i)-\ep_i)\exp(\ep_i-\exp(\ep_i))d\ep_i\nonumber\\
    &(\exp(\ep_i)=z)\nonumber\\
    &=\int^\infty_0(z\log(z)-\log(z))\exp(-z)dz\nonumber\\
    &=\int^\infty_0z\log(z)\exp(-z)dz-\int^\infty_0\log(z)\exp(-z)dz\nonumber\\
    &=[-z\log(z)\exp(-z)]^\infty_0+\int^\infty_0(\log(z)+1)\exp(-z)dz-\int^\infty_0\log(z)\exp(-z)dz\nonumber\\
    &=[-\exp(-z)]^\infty_0\nonumber\\
    &=1.
\end{align}
From Assumption 3, $\frac{\partial l(\bth)}{\partial\si}=O_p(\sqrt{n})$.

The second derivative of the likelihood function with respect to $\si$ is
\begin{align}
    \frac{\partial l(\bth)}{(\partial         \si)^2}=\sum^n_{i=1}\frac{\Det_i}{\Sh(Y_i)}\biggl(&\frac{1}{\si^2}-\frac{2}{\si^3}\X_i^T\bbe+\frac{2}{\si^3}\log(y_i)-\frac{y_i^\frac{1}{\si}\exp\left(-\frac{\X_i^T\bbe}{\si}\right)}{\si^4}(\log(y_i)-\X_i^T\bbe)^2\nonumber\\
    &\hspace{0.5cm}-\frac{2y_i^\frac{1}{\si}\exp\left(-\frac{\X_i^T\bbe}{\si}\right)}{\si^3}(\log(y_i)-\X_i^T\bbe)\biggr).
\end{align}
Because the sum of the constants is $n$, $\frac{\partial^2 l(\bth)}{(\partial \si^2)^2}=O_p(n)$.

The derivative of the likelihood function with respect to $\bbe$ and $\si^2$ is
\begin{align}
    \frac{\partial^2 l(\bth)}{\partial \si^2\partial \bbe}=\frac{1}{\si^2}\sum^n_{i=1}\frac{\Det_i}{\Sh(Y_i)}\left(1-\frac{y_i^\frac{1}{\si}\exp\left(-\frac{\X_i^T\bbe}{\si}\right)}{\si}(\log(y_i)-\X_i^T\bbe)-y_i^\frac{1}{\si}\exp\left(-\frac{\X_i^T\bbe}{\si}\right)\right)\X_i.
\end{align}
From Assumption 2, $\frac{\partial^2 l(\bth)}{\partial \si^2\partial \bbe}=O_p(\sqrt{n})$.

\subsection{Asymptotic properties of ISPW log-likelihood function of log-logistic distribution}
The ISPW log-likelihood function of the log-logistic distribution is 
\begin{align}
    l(\bth)=\sum^n_{i=1}\frac{\Det_i}{\Sh(Y_i)}\left(-\log(\si)-\frac{1}{\si}\X_i^T\bbe+\left(\frac{1}{\si}-1\right)\log(y_i)-2\log\left(1+y^{\frac{1}{\si}}\exp\left(-\frac{\X_i^T\bbe}{\si}\right)\right)\right).
\end{align}

The first derivative of the likelihood function with respect to $\bbe$ is
\begin{align}
    \frac{\partial l(\bth)}{\partial \bbe}=-\frac{1}{\si}\sum^n_{i=1}\frac{\Det_i}{\Sh(Y_i)}\left(1-2\frac{y^{\frac{1}{\si}}\exp\left(-\frac{\X_i^T\bbe}{\si}\right)}{1+y^{\frac{1}{\si}}\exp\left(-\frac{\X_i^T\bbe}{\si}\right)}\right)\X_i.
\end{align}
From Assumption 2, $\frac{\partial l(\bth)}{\partial \bbe\partial \bbe^T}=O_p(\sqrt{n})$.

The second derivative of the likelihood function with respect to $\bbe$ is
\begin{align}
    \frac{\partial l(\bth)}{\partial \bbe\partial \bbe^T}=-\frac{2}{\si^2}\sum^n_{i=1}\frac{\Det_i}{\Sh(Y_i)}\frac{y^{\frac{1}{\si}}\exp\left(-\frac{\X_i^T\bbe}{\si}\right)}{\left(1+y^{\frac{1}{\si}}\exp\left(-\frac{\X_i^T\bbe}{\si}\right)\right)^2}\X_i\X_i^T.
\end{align}
From Assumption 2, $\frac{\partial l(\bth)}{\partial \bbe\partial \bbe^T}=O_p(n)$.

The first derivative of the likelihood function with respect to $\si$ is
\begin{align}
    \frac{\partial l(\bth)}{\partial \si}=&\sum^n_{i=1}\frac{\Det_i}{\Sh(Y_i)}\left(-\frac{1}{\si}+\frac{1}{\si^2}\X_i^T\bbe-\frac{1}{\si^2}\log(y_i)+\frac{2}{\si^2}\frac{y^{\frac{1}{\si}}\exp\left(-\frac{\X_i^T\bbe}{\si}\right)}{1+y^{\frac{1}{\si}}\exp\left(-\frac{\X_i^T\bbe}{\si}\right)}(\log(y_i)-\X_i^T\bbe)\right)\nonumber\\
    =&\sum^n_{i=1}\frac{\Det_i}{\Sh(Y_i)}\left(-\frac{1}{\si}-\frac{\ep_i}{\si}+\frac{2}{\si}\frac{\ep_i\exp(\ep_i)}{1+\exp(\ep_i)}\right)\nonumber\\
    =&\sum^n_{i=1}\frac{\Det_i}{\Sh(Y_i)}\left(-\frac{1}{\si}+\frac{\ep_i}{\si}-\frac{2}{\si}\frac{\ep_i}{1+\exp(\ep_i)}\right).
\end{align}

The expectation of $\ep_i$ is
\begin{align}
    E[\ep_i]&=\int^\infty_{-\infty}\ep_i\frac{\exp(\ep_i)}{(1+\exp(\ep_i))^2}d\ep_i\nonumber\\
    &(\frac{1}{1+\exp(\ep_i)}=z)\nonumber\\
    &=\int^1_0(\log(1-z)-\log(z))dz\nonumber\\
    &(\int^1_0\log(1-z)dz=\int^1_0\log(z)dz)\nonumber\\
    &=0.
\end{align}

The expectation of $\frac{\ep_i}{1+\exp(\ep_i)}$ is
\begin{align}
    E\left[\frac{\ep_i}{1+\exp(\ep_i)}\right]&=\int^\infty_{-\infty}\frac{\ep_i}{1+\exp(\ep_i)}\frac{\exp(\ep_i)}{(1+\exp(\ep_i))^2}d\ep_i\nonumber\\
    &(\frac{1}{1+\exp(\ep_i)}=z)\nonumber\\
    &=\int^1_0z(\log(1-z)-\log(z))dz\nonumber\\
    &=\int^1_0z\log(1-z)dz-\int^1_0z\log(z)dz\nonumber\\
    &=\int^1_0\log(z)dz-2\int^1_0z\log(z)dz\nonumber\\
    &(\log(z)=\xi)\nonumber\\
    &=-[\exp(\xi)]^0_{-\infty}+2\left[\frac{1}{4}\exp(2\xi)\right]^0_{-\infty}\nonumber\\
    &=-\frac{1}{2}.
\end{align}
From Assumption 3, $\frac{\partial l(\bth)}{\partial\si}=O_p(\sqrt{n})$.

The second derivative of the likelihood function with respect to $\si$ is
\begin{align}
    \frac{\partial l(\bth)}{(\partial\si)^2}=\sum^n_{i=1}\frac{\Det_i}{\Sh(Y_i)}\biggl(&\frac{1}{\si^2}-\frac{2}{\si^3}\X_i^T\bbe+\frac{2}{\si^3}\log(y_i)-\frac{4}{\si^3}\frac{y^{\frac{1}{\si}}\exp\left(-\frac{\X_i^T\bbe}{\si}\right)}{1+y^{\frac{1}{\si}}\exp\left(-\frac{\X_i^T\bbe}{\si}\right)}(\log(y_i)-\X_i^T\bbe)\nonumber\\
    &\hspace{0.5cm}-\frac{2}{\si^4}\frac{\log(y_i)y^{\frac{1}{\si}}\exp\left(-\frac{\X_i^T\bbe}{\si}\right)}{1+y^{\frac{1}{\si}}\exp\left(-\frac{\X_i^T\bbe}{\si}\right)}(\log(y_i)-\X_i^T\bbe)\nonumber\\
    &\hspace{1cm}+\frac{2}{\si^4}\frac{\log(y_i)y^{\frac{1}{\si}}\exp\left(-\frac{\X_i^T\bbe}{\si}\right)}{\left(1+y^{\frac{1}{\si}}\exp\left(-\frac{\X_i^T\bbe}{\si}\right)\right)^2}(\log(y_i)-\X_i^T\bbe)\X_i^T\bbe\biggr).
\end{align}
Because the sum of the constants is $n$, $\frac{\partial^2 l(\bth)}{(\partial \si^2)^2}=O_p(n)$.

The derivative of the likelihood function with respect to $\bbe$ and $\si^2$ is
\begin{align}
    \frac{\partial^2 l(\bth)}{\partial \si^2\partial \bbe}=\sum^n_{i=1}\frac{\Det_i}{\Sh(Y_i)}\left(\frac{1}{\si^2}-\frac{2}{\si^3}\frac{y^{\frac{1}{\si}}\exp\left(-\frac{\X_i^T\bbe}{\si}\right)}{\left(1+y^{\frac{1}{\si}}\exp\left(-\frac{\X_i^T\bbe}{\si}\right)\right)^2}(\log(y_i)-\X_i^T\bbe)-\frac{2}{\si^2}\frac{y^{\frac{1}{\si}}\exp\left(-\frac{\X_i^T\bbe}{\si}\right)}{1+y^{\frac{1}{\si}}\exp\left(-\frac{\X_i^T\bbe}{\si}\right)}\right)\X_i.
\end{align}
From Assumption 2, $\frac{\partial^2 l(\bth)}{\partial \si^2\partial \bbe}=O_p(\sqrt{n})$.

\section{ISPW lasso with tuning parameter adjusted by cross-validation}
\subsection{Example}
The tuning parameter was set to 0.050 following a cross-validation study. The estimators of the intercept, age, and sex are $5.36$, $-0.03$, and $0.87$.

\subsection{Simulation results}
We present the MSEs in Table \ref{ATable1}.
\begin{table}[H]
  \begin{center}
\caption{Results of MSE \label{ATable1}}
\begin{tabular}{|c|c|c|}\hline
&Sample size=200&Sample size=1000\\\hline
\multicolumn{3}{|l|}{Scenario 1}\\\hline
$\be_0$& 0.021 & 0.006\\
$\be_1$& 0.079 & 0.058\\
$\be_2$& 0.059 & 0.054\\\hline
\multicolumn{3}{|l|}{Scenario 2}\\\hline
$\be_0$& 0.023 & 0.004\\
$\be_1$& 0.078 & 0.053\\
$\be_2$& 0.060 & 0.053\\\hline
\multicolumn{3}{|l|}{Scenario 3}\\\hline
$\be_0$& 0.051 & 0.038\\
$\be_1$& 0.145 & 0.122\\
$\be_2$& 0.119 & 0.113\\\hline
\multicolumn{3}{|l|}{Scenario 4}\\\hline
$\be_0$& 0.046 & 0.029\\
$\be_1$& 0.141 & 0.114\\
$\be_2$& 0.117 & 0.109\\\hline
\multicolumn{3}{|l|}{Scenario 5}\\\hline
$\be_0$& 0.107 & 0.077\\
$\be_1$& 0.205 & 0.151\\
$\be_2$& 0.156 & 0.140\\\hline
\multicolumn{3}{|l|}{Scenario 6}\\\hline
$\be_0$& 0.155 & 0.113\\
$\be_1$& 0.209 & 0.144\\
$\be_2$& 0.161 & 0.139\\\hline
\end{tabular}
\end{center}
\end{table}

The percentage of cases in which the correct variable combination was selected is given in Table \ref{ATable2}.
\begin{table}[H]
  \begin{center}
\caption{Percentage of times the correct variable combination is selected\label{ATable2}}
\begin{tabular}{|c|c|c|}\hline
&Sample size=200&Sample size=1000\\\hline
\multicolumn{3}{|l|}{Scenario 1}\\\hline
C1& 0.0 & 0.0\\
C2& 37.9 & 37.6\\
C3& 62.1 & 62.4\\
C4& 0.0 & 0.0\\\hline
\multicolumn{3}{|l|}{Scenario 2}\\\hline
C1& 0.0 & 0.0\\
C2& 37.2 & 37.9\\
C3& 62.7 & 62.1\\
C4& 0.0 & 0.0\\\hline
\multicolumn{3}{|l|}{Scenario 3}\\\hline
C1& 0.0 & 0.0\\
C2& 38.7 & 38.2\\
C3& 61.2 & 61.8\\
C4& 0.2 & 0.0\\\hline
\multicolumn{3}{|l|}{Scenario 4}\\\hline
C1& 0.0 & 0.0\\
C2& 39.0 & 38.0\\
C3& 60.7 & 62.0\\
C4& 0.3 & 0.0\\\hline
\multicolumn{3}{|l|}{Scenario 5}\\\hline
C1& 0.0 & 0.0\\
C2& 36.7 & 38.2\\
C3& 58.9 & 61.8\\
C4& 4.4 & 0.0\\\hline
\multicolumn{3}{|l|}{Scenario 6}\\\hline
C1& 0.0 & 0.0\\
C2& 36.4 & 38.4\\
C3& 57.8 & 61.6\\
C4& 5.7 & 0.0\\\hline
\end{tabular}
\end{center}
\end{table}

\subsection{Analysis of PBC data}
The tuning parameter was set to 0.206 following cross-validation. The estimators of the intercept and edema are $1.54$ and $-1.48$. Because of the large tuning parameter, most variables were removed.

\subsection{Analysis of MG data}
The tuning parameter was set to 0.011 following cross-validation. The estimators of the intercept, treatment group, age, years since diagnosis, Karnofsky performance score, race, radiation therapy, sex, nitro, path, and grade are 4.39, 0.14, $-0.01$, 0.02, 0.29, $-0.36$, 0.45, 0.03, $-0.36$, 0.06, and $-0.51$. No variables were removed because the tuning parameter was very small.

\section{Survival curves of each scenario in simulation study}
\begin{figure}[H]
  \begin{center}
  \includegraphics[width=15cm]{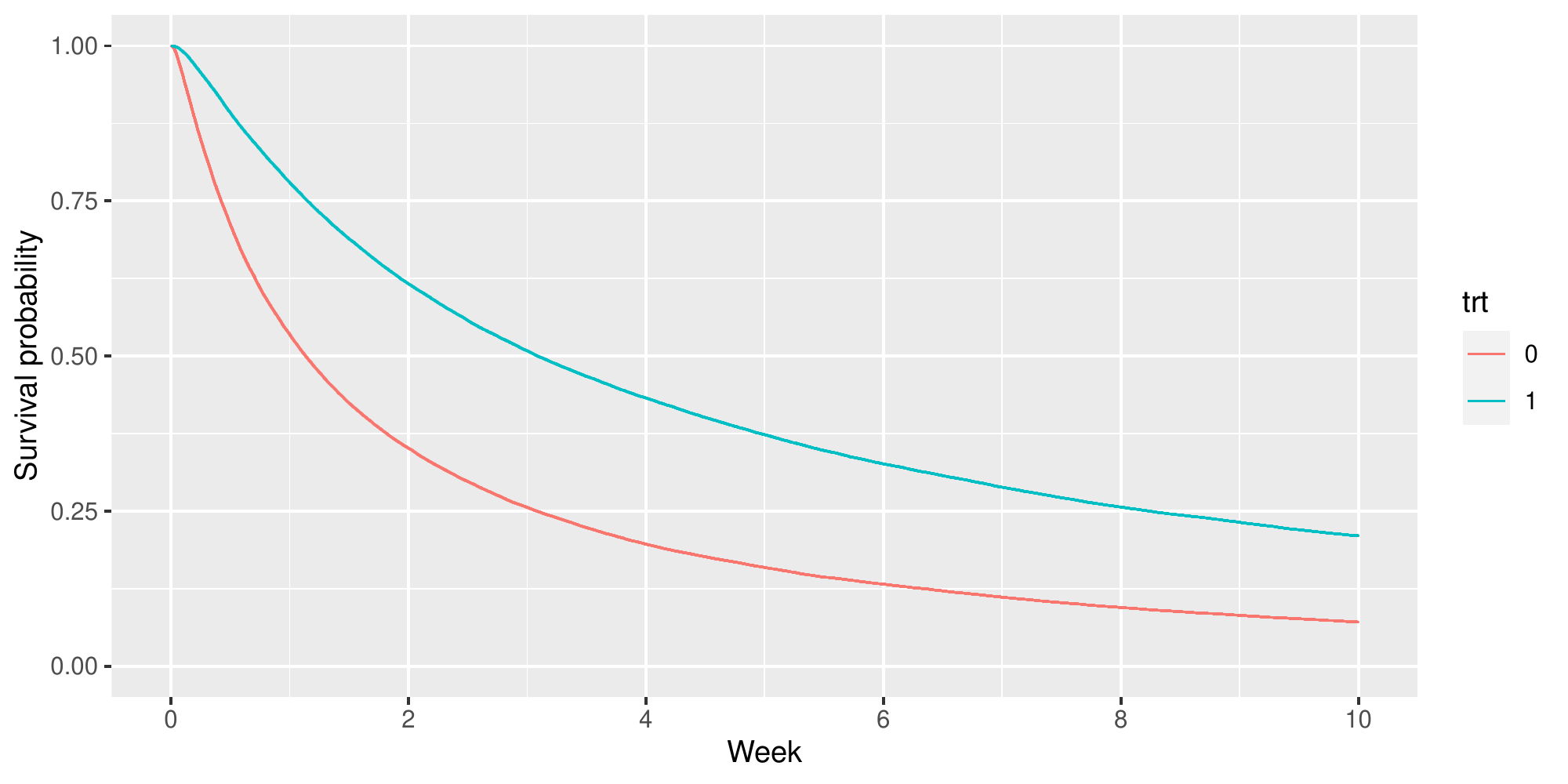}
  \caption{Survival probabilities of Scenario 1}
  \label{AFigure1}
  \end{center}
\end{figure}
\begin{figure}[H]
  \begin{center}
  \includegraphics[width=15cm]{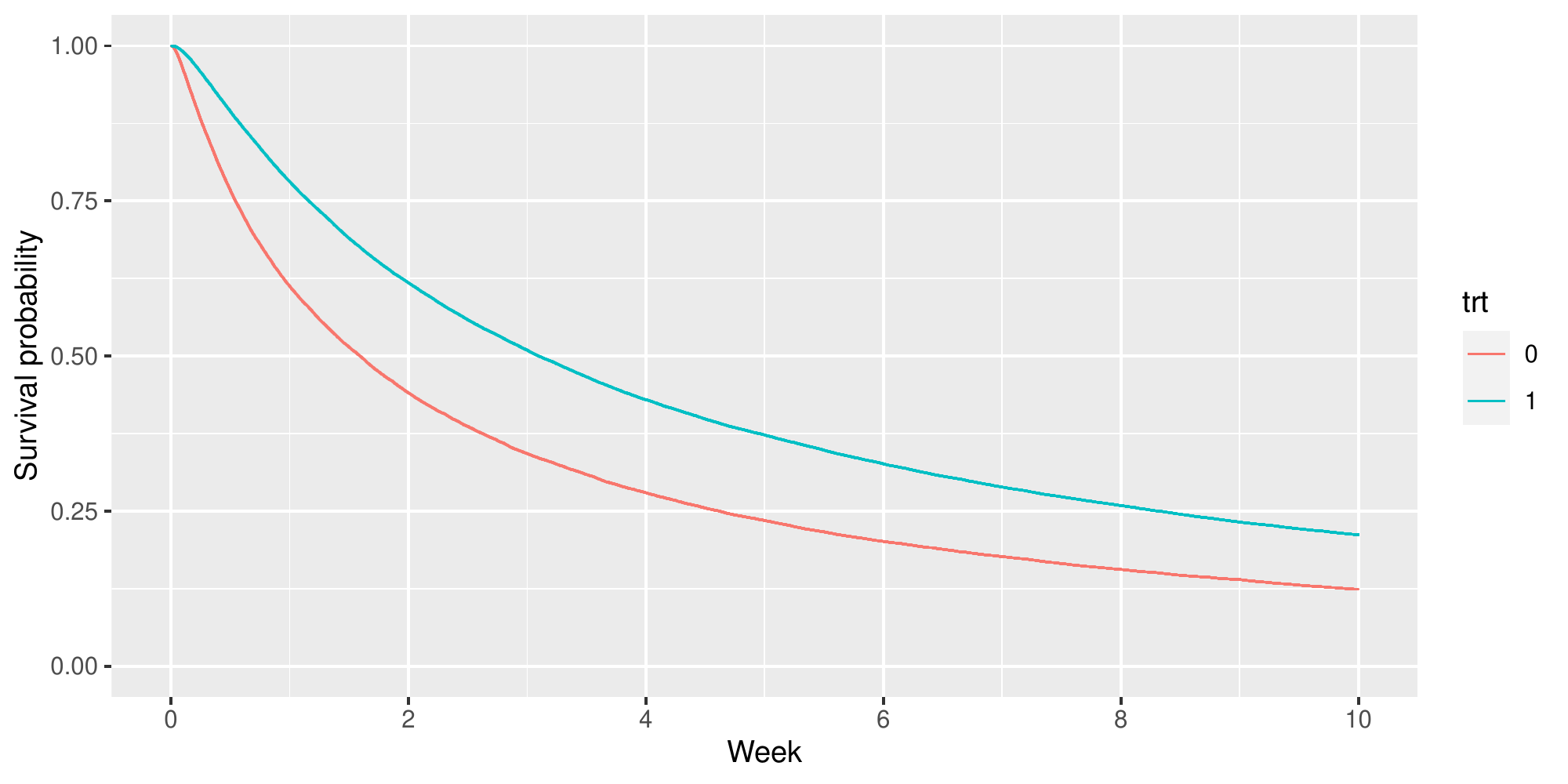}
  \caption{Survival probabilities of Scenario 2}
  \label{AFigure1}
  \end{center}
\end{figure}
\begin{figure}[H]
  \begin{center}
  \includegraphics[width=15cm]{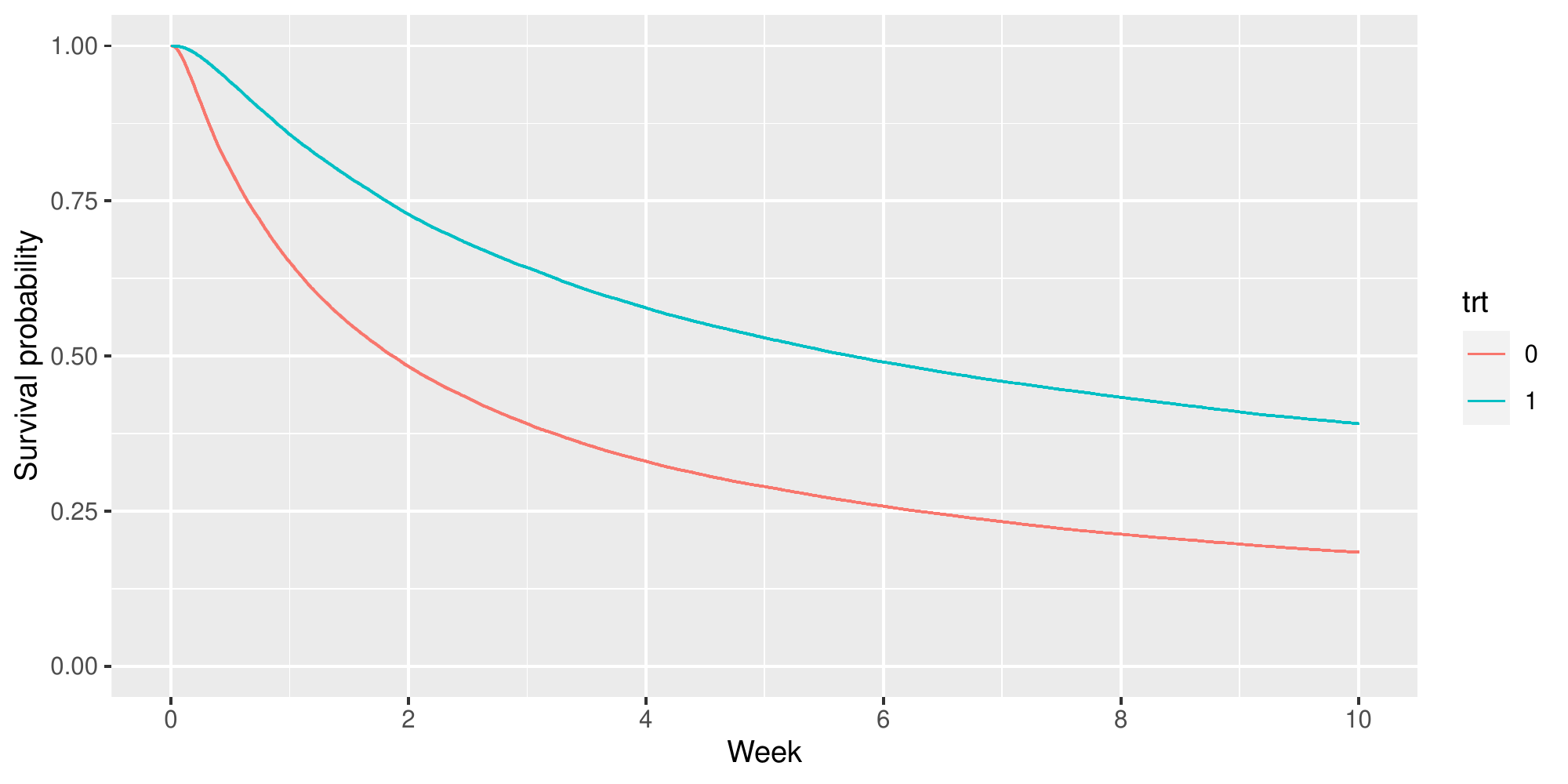}
  \caption{Survival probabilities of Scenario 3}
  \label{AFigure1}
  \end{center}
\end{figure}
\begin{figure}[H]
  \begin{center}
  \includegraphics[width=15cm]{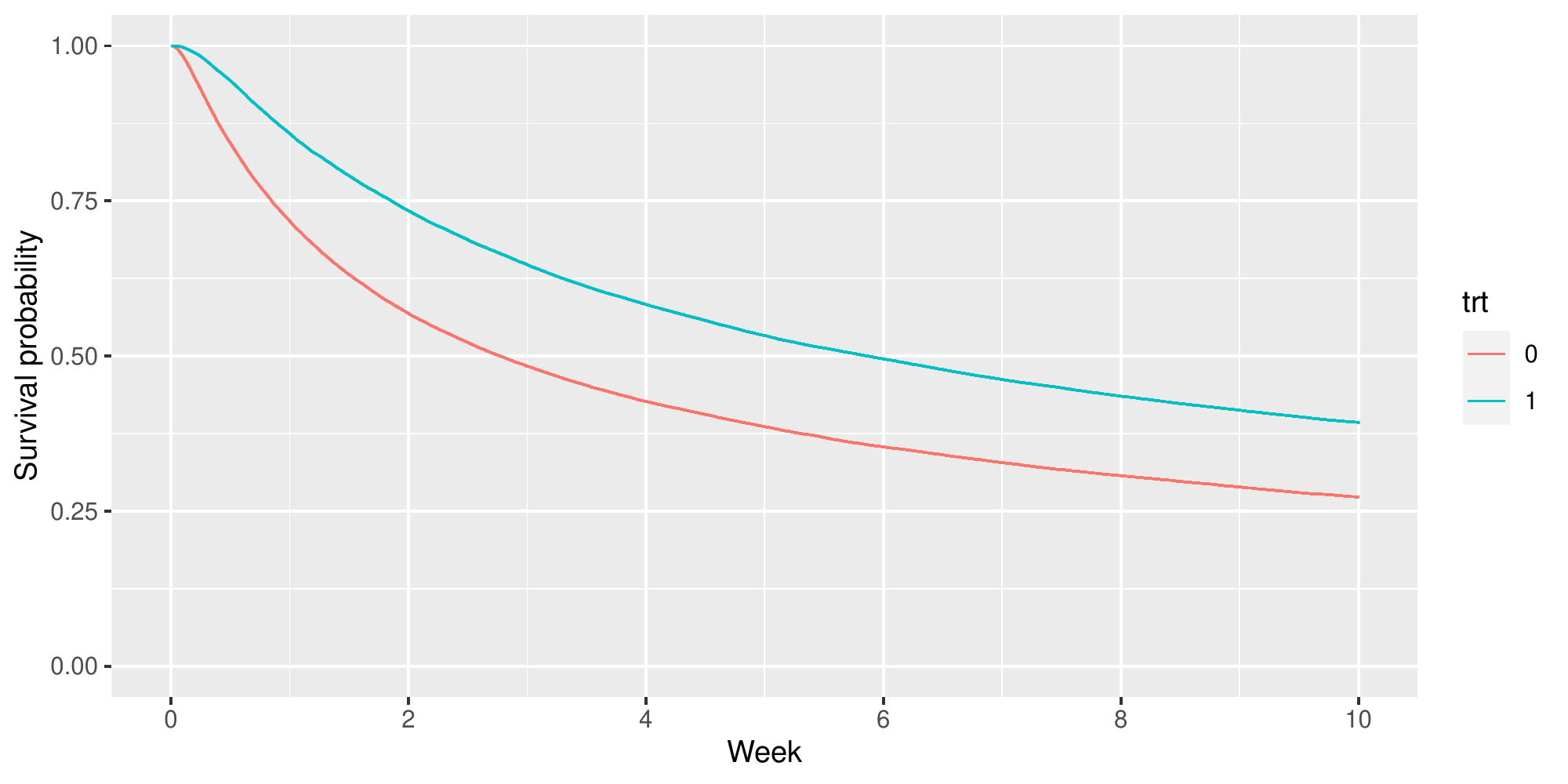}
  \caption{Survival probabilities of Scenario 4}
  \label{AFigure1}
  \end{center}
\end{figure}
\begin{figure}[H]
  \begin{center}
  \includegraphics[width=15cm]{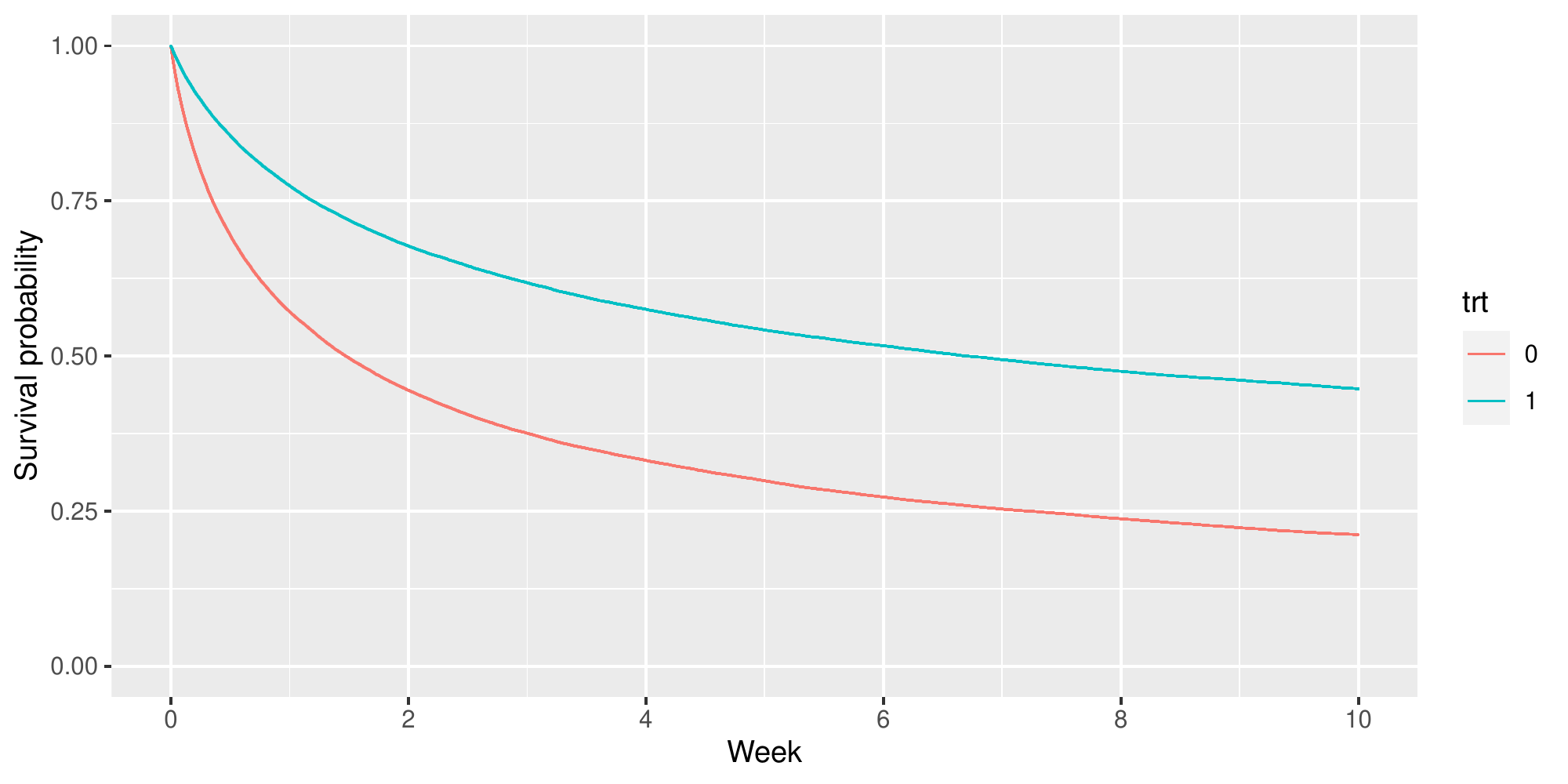}
  \caption{Survival probabilities of Scenario 5}
  \label{AFigure1}
  \end{center}
\end{figure}
\begin{figure}[H]
  \begin{center}
  \includegraphics[width=15cm]{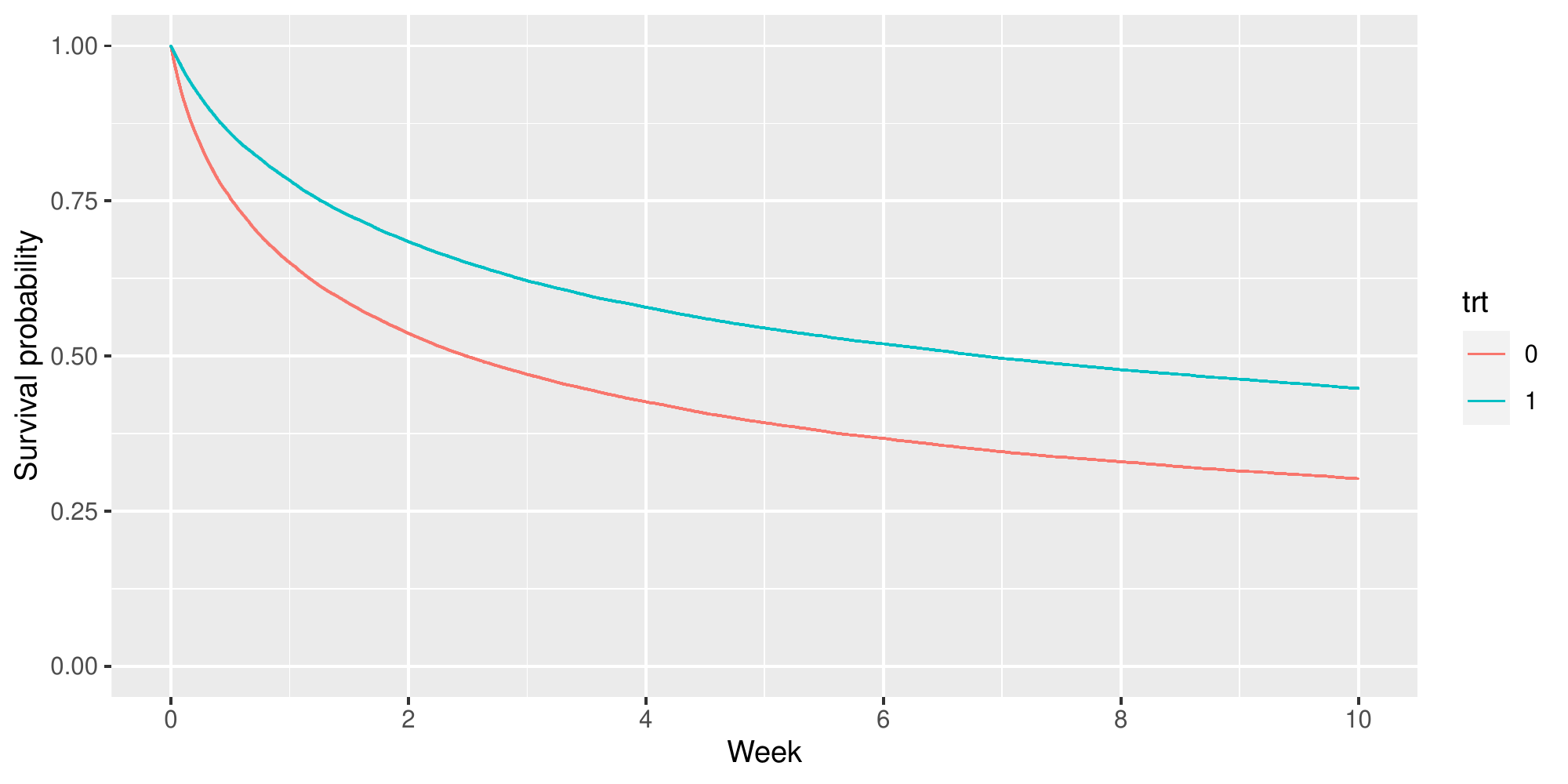}
  \caption{Survival probabilities of Scenario 6}
  \label{AFigure1}
  \end{center}
\end{figure}

\end{document}